\documentclass[a4paper,reqno]{amsart}
\usepackage[text={5.2in,8in},centering]{geometry}
\usepackage[backrefs]{amsrefs}
\usepackage{amssymb}
\usepackage{enumerate}
\usepackage{graphicx}
\usepackage{color}

\definecolor{trustcolor}{rgb}{0.71,0.14,0.07}

\numberwithin{equation}{section}
\theoremstyle{plain}
\newtheorem{theorem}{Theorem}[section]
\newtheorem{prop}{Proposition}

\newtheorem{cor}{Corollary}
\newtheorem{lemma}{Lemma}[section]
\theoremstyle{remark}
\newtheorem{remark}{Remark}[section]

\newtheorem*{quest*}{Question}
\newtheorem*{remark*}{Remark}
\theoremstyle{remark}

\theoremstyle{definition}
\newtheorem{definition}{Definition}[section]
\newtheorem*{definition*}{Definition}
\newtheorem*{notation*}{Notation}
\newtheorem*{notations*}{Notations}
\providecommand{\B}{\mathbf}
\providecommand{\BS}[1]{\boldsymbol{#1}}
\providecommand{\C}{\mathcal}
\providecommand{\D}{\mathbb}

\newcommand{\ee}{\mathrm{e}}

\DeclareMathOperator{\dist}{dist}
\DeclareMathOperator{\card}{\#}

\DeclareMathOperator{\one}{\mathbf{1}}

\DeclareMathOperator{\supp}{supp}

\DeclareMathOperator{\diam}{{\rm diam}}

\def\ux{\B{x}}
\def\uy{\B{y}}
\def\uu{\B{u}}
\def\uv{\B{v}}

\def\uzero{\B{0}}

\def\uA{\B{A}}

\def\uC{\B{C}}
\def\uG{\B{G}}
\def\uH{\B{H}}

\def\uU{\B{U}}
\def\uV{\B{V}}

\def\uDelta{\B{\Delta}}

\def\uPsi{\B{\Psi}}
\def\uPhi{{\BS{\Phi}}}

\def\uw{\B{w}}
\def\DC{\D{C}}
\def\DD{\D{D}}
\def\DR{\D{R}}
\def\DZ{\D{Z}}
\def\DN{\D{N}}

\def\cA{\C{A}}

\def\cE{\C{E}}

\def\cR{\C{R}}
\def\cS{\C{S}}

\def\om{{\omega}}

\def\CMnu{\B{CM(\nu)}}
\def\DSk{{$\B{(DS.I,k,N)}$}}
\def\DSkone{{$\B{(DS.I,k+1,N)}$}}
\def\DS#1{{\B{(DS.I,#1)}}}

\def\half{{\frac{1}{2}}}

\def\tr{{ {\rm tr\,}}}

\def\fF{\mathfrak{F}}
\def\fS{\mathfrak{S}}

\def\prob#1{\D{P}\left\{\,#1\,\right\}}
\def\esm#1{\D{E}\left[\,#1\,\right]}

\def\tto#1{\smash{\mathop{\,\,\,\, \longrightarrow \,\,\,\, }\limits_{#1}}}

\def\be{\begin{equation}}
\def\ee{\end{equation}}
\def\ba{\begin{array}{l}}
\def\ea{\end{array}}
\def\diy{\displaystyle}

\begin{document}
\title[Resonances and localization in multi-particle systems]
{Resonances and localization \\ in  multi-particle disordered systems}

\author[V. Chulaevsky]{Victor Chulaevsky}


\address{D\'{e}partement de Math\'{e}matiques\\
Universit\'{e} de Reims, Moulin de la Housse, B.P. 1039\\
51687 Reims Cedex 2, France\\
E-mail: victor.tchoulaevski@univ-reims.fr}

\begin{abstract}
This is a complement to our earlier work \cite{C10a} where a new eigenvalue concentration bound
for multi-particle disordered quantum lattice systems was obtained. Here we show that the new bound leads to a simplified proof of multi-particle spectral and dynamical localization.

\end{abstract}
\maketitle
%
\section{Introduction. The model and the motivation for this paper} \label{sec:intro}

We study multi-particle quantum systems in a disordered environment, usually referred to as Anderson-type models.
Consider an $N$-particle tight-binding Hamiltonian $\uH_{V,\uU}(\omega)$ in the Hilbert space $\ell^2(\DZ^{Nd})$,
\begin{equation}\label{eq:H}
\uH_{V,U} = \uDelta + g\uV + \uU  = \sum_{j=1}^N \left( \Delta^{(j)} + gV(x_j, \omega)\right) + \uU,
\end{equation}
where $\uDelta$ is the nearest-neigbor lattice Laplacian,
$$
\Delta^{(j)}  \Psi(\ux) \equiv \Delta^{(j)}  \Psi(x_1, \ldots, x_N) = \sum_{y\in\DZ^d:\, |y|=1}  \Psi(x_1, \ldots, x_j + y,  \ldots, x_N),
$$
$V:\DZ^d\times \Omega \to \DR$ is a random field relative to a probability space
$(\Omega, \C{F}, \D{P})$, $g>0$ is a constant measuring the "amplitude" of the potential $V$, and $\uU$ is the multiplication operator by a function $\uU(\ux)$ which we assume bounded, but not necessarily symmetric. In addition,  we  assume that $\uU$ has a finite range $r_0<\infty$. In the case of a two-body interaction generated by an interaction potential
$U^{(2)}:\DZ^{d} \to \DR$,
$$
\uU(x_1, \ldots, x_N) = \sum_{i<j} U^{(2)}(|x_i - x_j|),
$$
this hypothesis is equivalent to the following condition:  $\supp U^{(2)}\subset [0,r_0]$.
The assumptions on the random field $V$ are described in Sect.~\ref{ssec:main.res}.
Unless otherwise specified, the boldface symbols denote objects relative to multi-particle systems.

Given any finite cube $\uC_{L}(\uu) := \{\ux\in\DZ^{Nd}:\, \|\ux - \uu\| \le L\}$,
we will consider a finite-volume approximation of the Hamiltonian $\uH$
$$
\uH_{\uC_{L}(\uu)} = \uH \upharpoonright_{\ell^2(\uC_{L}(\uu))} \text{ with Dirichlet boundary conditions on } \partial \uC_{L}(\uu)
$$
acting in the finite-dimensional Hilbert space $\ell^2(\uC_{L}(\uu))$.
In \cite{CS08} the following "two-volume" version of the Wegner bound was established for pairs of two-particle operators $\uH_{\uC_{L}(\uu)}$, $\uH_{\uC_{L'}(\uu')}$  such that $L \ge L'$ and
$\dist(\uC_{L}(\uu), \uC_{L'}(\uu') ) \ge 8L$: if $\nu$ is the continuity modulus of the marginal distribution function $F_V$, then
$$
\prob{ \dist( \sigma(\uH_{\uC_{L}(\uu)} ), \sigma( \uH_{\uC_{L'}(\uu')}) \le \epsilon}\\
\le \, (2L+1)^{2d} (2L'+1)^d \,\nu(2\epsilon).
\eqno({\rm W2})
$$

\subsection {More efficient eigenvalue concentration bounds}\label{ssec:motiv}

In \cite{CS09a,CS09b} the MPMSA was used to prove \textit{spectral} localization (i.e., exponential decay of eigenfunctions) in the strong disorder regime. Aizenman and Warzel \cite{AW09a,AW09b} used the FMM to prove directly \textit{dynamical} localization (hence, spectral localization) in various regions in the parameter space, including strong disorder, "extreme" energies and weak interactions.

However, due to a highly correlated nature of the potential energy in multi-particle systems, it was difficult to obtain optimal decay bounds for eigenfunctions in terms of some norm in $\DZ^{Nd}$. This difficulty  had been analyzed by Aizenman and Warzel \cite{AW09a} and served as the motivation for a more efficient eigenvalue concentration bound proven in \cite{C10a} (cf. Thm.~\ref{thm:W2.general} below). Here we prove spectral and dynamical localization for multi-particle systems with the potential $V(x;\om)$ satisfying the hypotheses of Thm.~\ref{thm:W2.general}, for $|g|$ large enough
\footnote{An adaptation of our method to the case of weak disorder, at "extreme" energies, is the subject of a fortgcoming manuscript by T. Ekanga \cite{E10}.}
 (strong disorder regime).

We would like to emphasize that the novelty of the present paper consists in the observation that the proof of dynamical (and spectral) localization for multi-particle systems, based upon Thm.~ \ref{thm:W2.general}, can be obtained by a \emph{minor}  modification of the single-particle version, once the key MSA bounds are established for the multi-particle system in question. In the author's opinion, this opens a way to numerous extensions of existing "single-particle" techniques to disordered systems with interaction.

\subsection{Basic geometrical definitions}

Consider  the lattice $(\DZ^{d})^N \cong \DZ^{Nd}$, $N>1$. Vectors
$\ux=(x_1, \ldots, x_N)\in\DZ^{Nd}$ will be identified with  $N$-particle configurations in $\DZ^d$. We use below the max-norm  $\|\cdot\|_\infty$ in $\DR^{nd} \supset \DZ^{nd}$, $n\ge 1$:
$$
 \|\ux \|_\infty  = \| (x_1, \ldots, x_n) \|_\infty
 := \max_{j\in[[1, n]]}  \max_{i\in[[1, d]]} |x_j^{(i)}|.
$$
This norm canonically induces the notion of diameter for subsets of $\DR^{nd}$ and $\DZ^{nd}$, denoted below as "$\diam$".
We denote by $\DD$ the "principal diagonal" in $(\DZ^{d})^N $:
$$
\DD = \{ \ux\in\DZ^{Nd}: \,\ux = (x, \ldots, x), \, x\in\DZ^d \}.
$$
We will often use the standard notation $[[a,b]] := [a,b]\cap \DZ$.

Taking into account the symmetry of the  potential energy $V(x_1;\om) + \cdots + V(x_N;\om)$, it is natural to introduce also  the "symmetrized" distance
$$
d_S(\ux, \uy) := \min_{\tau\in \fS_N} \|\ux - \uy\|.
$$
Given a cube $\uC_L(\uu) = \{\ux:\, \|\ux - \uu\|\le L\} \subset \DZ^{Nd}$, we define three kinds of its "boundaries":
$$
\ba
\diy \partial^- \uC_L(\uu) = \{\ux:\, \|\ux - \uu\| = L\},  \\
\diy \partial^+ \uC_L(\uu) = \{\ux:\, \|\ux - \uu\| = L+1 \}, \\
\diy \partial \uC_L(\uu) =
\{(\ux, \ux'):\, \ux\in\partial^- \uC_L(\uu), \ux'\in\partial^+ \uC_L(\uu),
\|\ux - \ux'\| = 1 \}.
\ea
$$

\subsection{ The main result on multi-particle eigenvalue concentration}\label{ssec:main.res}

Introduce the following notations. Given a parallelepiped  $Q\subset \DZ^d$, we denote by $\xi_{Q}(\omega)$ the sample mean of the random field $V$ over  $Q$,
$$
 \xi_{Q}(\omega) =   \frac{1}{| Q |} \sum_{x\in Q} V(x,\omega)
$$
and introduce the "fluctuations" of $V$ relative to the sample mean,
$$
 \eta_x  = V(x,\omega) - \xi_{Q}(\omega), \; x\in Q.
$$
We denote by $\fF_{V, Q}$ the sigma-algebra generated by
$\{\eta_x, \,x\in Q; V(y;\cdot), y\not\in Q\}$, and by
$F_\xi( \cdot\,| \fF_{V, Q})$ the conditional distribution function of $\xi_Q$ given
$\fF_{V, Q}$:
$$
 F_\xi(s\,| \fF_{V, Q}) := \prob{ \xi_Q \le s\,|\, \fF_{V, Q} }.
$$

We will  assume that the random field $V$ satisfies the following condition:
\par\medskip
($\CMnu$):
\textit{ For any $R \ge 0$ there exists a function $\nu_R:\DR_+\to\DR_+$ vanishing at $0$ and such that $\forall\, Q\subset  \DZ^d$ with $\diam(Q)\le R$ the conditional distribution  function
$F_\xi( \cdot \,| \fF_{V, Q})$ satisfies
\begin{equation}\label{eq:CMnu}
\forall \, t,s\in\DR, \;\;
{\rm ess} \sup
|F_\xi(t\,| \fF_{V, Q}) - F_\xi(s\,| \fF_{V, Q})| \le \nu_R(|t-s|).
\end{equation}
} 

\smallskip

This condition may prove useless for applications if $\nu_R(s)\downarrow 0$  too slowly as $s\downarrow 0$.  For this reason, we assume below, in addition to \eqref{eq:CMnu}, that
\begin{equation}\label{eq:CMnu.stronger}
\nu_R(t) \le Const \, R^{A} \ln^{-B} |t|^{-1}, \;\; |t| < 1,
\end{equation}
for some $A<\infty$ and sufficiently large $B$. Alternatively, a stronger condition of H\"{o}lder- or Lipshitz-continuity can be used: for some  $A<\infty$ and $b>0$,
\begin{equation}\label{eq:CMnu.stronger.Holder}
\nu_R(t) \le Const \, R^{A} |t|^{b}, \;\; |t| < 1.
\end{equation}

\smallskip

Note that in the particular case of a Gaussian  IID field $V$  with zero mean and unit variance, $\xi_Q$ is a Gaussian random variable with variance ${|Q|}^{-1}$, independent of the "fluctuations" $\eta_x$, so that its probability density exists and is bounded:
$$
p_{\xi_{Q}}(s) = |Q|^{1/2} \, (2\pi)^{-1/2}\, e^{-\frac{{|Q|} s^2}{2 } } \le |Q|^{1/2} \, (2\pi)^{-1/2},
$$
although  $\|p_{\xi_{Q}}\|_\infty$  grows with $|Q|$, and so does the continuity modulus of $F_{\xi_{Q}}$.

\begin{theorem}[Cf. \cite{C10a}]\label{thm:W2.general}
Let $V: \DZ^d\times \Omega \to \DR$ be a random field satisfying
{\rm \textbf{($\CMnu$)}}.
Then for any pair of $N$-particle operators $\uH_{\uC_{L'}(\uu')}$, $\uH_{\uC_{L''}(\uu'')}$,
$0 \le L', L'' \le L$, satisfying $d_S(\uu', \uu'') > 2NL$, and any $s>0$ the following bound holds:
\begin{equation}\label{eq:main.W2.bound}
\prob{ \dist(\sigma(\uH_{\uC_{L'}(\uu')}), \sigma(\uH_{\uC_{L''}(\uu'')})) \le s }
     \le \displaystyle|\uC_{L'}(\uu')| \cdot |\uC_{L''}(\uu'')|\, \nu_L (2s).
\end{equation}
\end{theorem}

An adaptation of our method to a class of bounded  potentials $V(\cdot;\omega)$, for which $(\CMnu)$ does not hold, is the subject of a forthcoming manuscript by M. Gaume \cite{G10}.

\section{Multi-particle MSA}

\subsection{Decay of resolvents in finite cubes: the key bound}

\begin{definition}\label{def:S}
Given a sample $\uH(\om)$ of an $N$-particle Hamiltonian of the form \eqref{eq:H}, a cube $\uC_L(\uu)$ is said $(E,m)$-non-singular ($(E,m)$-NS), with $E\in \DR$ and $m>0$, if
\begin{equation}\label{eq:def.NS}
\max_{\ux:\, \| \ux - \uu \|\le L^{1/\alpha} }\, \max_{y\in \partial^- \uC_{L}(\uu)}
\; | (\uH_{\uC_L(\uu)} - E)^{-1}(\ux, \uy)| \le e^{-\gamma(m,L,N)},
\end{equation}
where
$$
\gamma(m,L,n) := mL(1+L^{-1/4})^{N-n+1}, \;\; 1 \le n \le N.
$$
Otherwise, it is called $(E,m)$-singular ($(E,m)$-S).
\end{definition}

Traditionally,  the $k$-dependence is put in the decay exponent, $m=m_k$, which is re-calculated recursively. We prefer to keep fixed the  decay parameter $m$, while the actual decay bound depends on $L_k$ and on $N$ explicitly, through the function $\gamma$.

The main result of the multi-particle MSA, used for the derivation of spectral and dynamical localization, can be formulated as follows:
\smallskip

$\DS{k, n}$: \quad \textit{ For any pair of $2nL_k$-distant $n$-particle cubes $\uC_{L_k}(\ux)$, $\uC_{L_k}(\uy)$ the following bound holds:}
\begin{equation}\label{eq:DS}
\begin{array}{ll}
\prob{\exists\, E\in I:\,\, \text{ $\uC_{L_k}(\ux)$ and $\uC_{L_k}(\uy)$ are $(E,m)$-S  } }
\le L_k^{-p\,2^{N-n+1}}.
\end{array}
\end{equation}

\subsection{Geometric resolvent inequalities}

The second  resolvent identity implies  the following formula for the resolvent $\uG_{\uC_L(\uu)}(\ux,\uy; \zeta)$ of  the lattice Schr\"{o}dinger operator in a cube
$\uC_L(\uu) \supset \uC_\ell(\uw)$, $L > \ell+1$: for any complex $\zeta$ which is not an eigenvalue of operators $\uH_{\uC_L(\uu)}$ and $\uH_{\uC_\ell(\uu)}$, and
$\ux\in\uC_\ell(\uw)$, $\uy\in\uC_L(\uu) \setminus \uC_\ell(\uw)$,
\be
\uG_{\uC_L(\uu)}(\ux,\uy; \zeta)
= \sum_{(\uv,\uv')\in\partial \uC_\ell(\uw)}
\uG_{\uC_\ell(\uw)}(\ux,\uv; \zeta) \uG_{\uC_L(\uu)}(\uv',\uy; \zeta)
\ee
yielding the well-known Geometric Resolvent Inequality (GRI):
\be\label{eq:GRI.GF}
\ba
\diy \left| \uG_{\uC_L(\uu)}(\ux,\uy; \zeta) \right|
\le |\partial \uC_\ell(\uw)|\, \max_{(\uv,\uv')\in\partial \uC_\ell(\uw)}
 |\uG_{\uC_\ell(\uw)}(\ux,\uv; \zeta)| \,  |\uG_{\uC_L(\uu)}(\uv',\uy; \zeta)| \\
\diy \le |\partial \uC_\ell(\uw)|\,\left(\max_{\uv\in\partial^- \uC_\ell(\uw)}  |\uG_{\uC_\ell(\uw)}(\ux,\uv; \zeta)|\right) \,
\max_{\uv'\in\partial^+ \uC_\ell(\uw)} |\uG_{\uC_L(\uu)}(\uv',\uy; \zeta)|.
\ea
\ee
Furthermore, if $\uPsi_n$ is an eigenfunction of the operator $\uH_{\uC_L(\uu)}$, then for any complex $\zeta$ which is not an eigenvalue of operator  $\uH_{\uC_\ell(\uu)}$ the following identity holds true (we will call it the Geometric Resolvent Inequality for eigenfunctions):
\be
\uPsi_n(\ux) = \sum_{(\uv,\uv')\in\partial \uC_\ell(\uw)}
\uG_{\uC_\ell(\uw)}(\ux,\uv; \zeta) \uPsi_n(\uv'),
\;\;\; \ux \in\uC_\ell(\uw),
\ee
yielding the GRI for eigenfunctions:
\be\label{eq:GRI.EF}
\ba
\diy \left| \uPsi_n(\ux) \right|
 \le |\partial \uC_\ell(\uw)|\,\left(\max_{\uv\in\partial^- \uC_\ell(\uw)}  |\uG_{\uC_\ell(\uw)}(\ux,\uv; \zeta)|\right) \, \max_{\uv'\in\partial^+ \uC_\ell(\uw)} | \uPsi_n(\uv') |.
\ea
\ee
The derivation of the GRI can be found, e.g., in the review \cite{K07}.

\subsection{Base of the induction in $N$}
\label{ssec:base.MSA}

In the course of the induction in $n$, we will assume that \eqref{eq:DS} is established for all $n\le N-1$ and all $k\ge 0$, and then prove it for $n=N$, at each scale $L_k$. The parameter $p>0$ will be required to be sufficiently large (cf. \eqref{eq:cond.p}), and the exponent $2^{N-n+1}p$ in the RHS of \eqref{eq:DS} shows that the strongest bound is required for the single-particle systems ($n=1$). It follows from the results of single-particle MSA that $p$ can be made arbitrarily large, if the disorder parameter $g$ is large enough. The single-particle FMM provides even an exponential bound of the LHS in \eqref{eq:DS}  for $n=1$.  See, e.g., the articles \cite{FMSS85,DK89,AM93}, the review \cite{K07}  and the monograph \cite{St01}.
As $n$ grows, the upper bound in \eqref{eq:DS} becomes weaker, hence, simpler to prove.

\subsection{Base of the scale induction, for a fixed $N$}

\begin{lemma}\label{lem:L0.MSA}
There exists $g^*<\infty$ and positive functions $m^*(g), p^*(g)$ defined for $|g|\ge g^*$ such that
\begin{enumerate}
  \item $m^*(g), p^*(g) \to + \infty $ as $|g|\to \infty$;
  \item the property $\DS{0,N}$ holds for the operators $\uH^{(N)}(\om) = \uDelta^{(N)} + g \uV(\omega) + \uU$  with $m=m^*(g)$ and $p=p^*(g)$.
\end{enumerate}
\end{lemma}

This statement can be proven for $N$-particle Hamiltonians in the same way as for $N=1$; see, e.g., \cite{FMSS85,DK89}. In fact, the non-random interaction $\uU$ (and/or a non-random component of $V$) can simply be \emph{ignored in the proof}. Note for a reader not familiar with the conventional MSA techniques that $m^*(g) \sim O(\ln |g|)$, for $|g|\gg 1$. The asymptotic behavior of $p^*(g)$ depends upon the regularity of the marginal distribution function $F_V$ of the random field $V$.

\subsection{"Radial descent" bound for resolvents}

In this subsection we state an analytic result which does not rely upon single- or multi-particle structure or random nature of the potential energy of the Hamiltonian $\uH$. All constants involved depend only upon the dimension of the lattice. \emph{In order to emphasize this, we change here our notations and do not use boldface symbols.} Since $V$ is arbitrary here, in applications to the multi-particle systems we can assume that $V$ contains also an inter-particle interaction energy. In a certain sense, Lemma \ref{lem:radial.descent} below encapsulates an argument going back to \cite{DK89,FMSS85} and used since then in many papers.

\begin{definition}\label{def:subharm}
A bounded function $f:\Lambda \to \DC$ on a subset $\Lambda\subset\DZ^n$, $n\ge 1$, is called $(\ell,q,\cS,c)$-subharmonic, with $\ell\in\DN$, $q>0$, $\cS \subset \Lambda$, $c\ge 1$, if for any
$x\in\Lambda\setminus\cS$ such that $C_{\ell}(x)\subset\Lambda$, one has
$$
|f(x)| \le q \max_{\|y-x\|=\ell} |f(y)|,
$$
while for $x\in\cS$
$$
|f(x)| \le q \,\max_{y\in\Lambda: \ell \le \|y-x\|\le (1+c)\ell} \,|f(y)|.
$$

\end{definition}

\begin{lemma}\label{lem:radial.descent}
Consider an $(\ell,q,\cS,c)$-subharmonic function $f$ on $C_L(u)\subset\DZ^n$. Suppose that the $c\ell$-neighborhood of the set $\cS$ can be covered by a collection $\cA$ of annuli
$$
A_i = C_{b_i}(u)\setminus C_{a_i}(u),
$$
of total width $W_{\cA}$.
Then for any $r \in[W_{\cA} +\ell, L - W_{\cA} +\ell]$
$$
\max_{x\in C_r(u)}
|f(x)| \le q^{[(L-r - W(A))/\ell] - 1} \max_{y\in C_L(u)} |f(y)|.
$$
In particular,
$$
|f(0)| \le q^{[(L- W(A))/\ell] - 1} \max_{y\in C_L(u)} |f(y)|.
$$
\end{lemma}

The proof,  based on a "reverse induction" in $r$ ("radial descent") is given in Sect.~\ref{sec:RDL.proof}. A direct application of this bound to the resolvents, making use of the GRI,  leads to the following statement.

\begin{lemma}\label{lem:radial.descent.GF}
Consider a lattice Schr\"{o}dinger operator $\Delta + V(x)$ in a cube $C_{L_{k+1}}(u)\subset \DZ^{n}$, $n \ge 1$.
Fix an energy $E\in I$ and suppose that the $cL_k$-neighborhood of all $(E,m)$-singular cubes of radius $L_k$ inside $C_{L_{k+1}}(u)$ can be covered by a collection $\cA$ of annuli
$A_i = C_{b_i}(u)\setminus C_{a_i}(u)$ of total width $W_{\cA}$.
Suppose also that $C_{L_{k+1}}(u)$  does not contain any $E$-resonant cube of radius $L \ge L_k$ (including itself). Then, for any fixed value of the constant $\widetilde{c}$ and $L_0 \ge L_0^*(c,n)$ large enough, the cube $C_{L_{k+1}}(u)$ is $(E,m)$-NS.
\end{lemma}

We will see below (cf. Sect.~\ref{ssec:FI.pairs}) that,  in application to MPMSA, with sufficiently high  probability a cube of radius $L_{k+1}$ does not contain more than $4$ singular cubes of radius $L_k$ which are pairwise $2NL_k$-distant, in which case all singular cubes can be covered  by a collection of annuli of total width $O(L_k)$.

We stress that Lemma \ref{lem:radial.descent.GF} is purely "deterministic" and does nor rely upon single- or multi-particle structure of the potential.

\subsection{Localization bounds for decomposable systems}

We need a simple property of quantum systems decomposed in a union of  perfectly non-interacting subsystems. Loosely speaking, it says that if two distant subsystems $\uu'\in\DZ^{n'}$, $\uu''\in\DZ^{n''}$ are "localized" and do not interact, then their union $\uu = (\uu', \uu'')$ is also "localized".

\begin{definition}\label{DefBad}
\par$\,$
\begin{enumerate}
  \item Let  $n'\in \{1, \ldots, N-1\}$, $k\ge 0$ and $\uu'=(u_1, \ldots, u_{n'})
\in\DZ^{n'd}$. Given a bounded interval $I\subset \DR$ and $m>0$, the $n'$-particle cube
$\uC^{(n')}_{L_k}(\uu')$ is said $(m,I)$-tunneling ($(m,I)${\rm{-T}}, for short) if
$
\exists\, E\in I$  and there are $2NL_{k-1}$-distant $n$-particle cubes $\uC^{(n')}_{L_{k-1}}(\uv_1),
\uC^{(n')}_{L_{k-1}}(\uv_2) \subset \uC^{(n')}_{L_{k}}(\uu')$  which are $(E,m)${\rm-S}.
  \item An $N$-particle cube $\uC^{(N)}_{L_k}(\uu)$ is said $(m,I)$-partially tunneling ($(m,I)${\rm-PT}) if, for some permutation $\tau \in \mathfrak{S}_N$ (acting on the components of the vectors $\uu = (u_1, \ldots, u_N)$) and some $n', n''\geq 1$, it admits a representation
$$
\uC^{(N)}_{L_k}(\tau(\uu) ) = \uC^{(n')}_{L_{k-1}}( \uu') \times \uC^{(n'')}_{L_{k-1}}( \uu'')
$$
$\uu' =(u_1, \ldots, u_{n'})$, $\uu'' =(u_{n'+1}, \ldots, u_{N})$, where
either
$\uC^{(n')}_{L_{k-1}}( \uu')$ or $\uC^{(n'')}_{L_{k-1}}( \uu'')$ is $(m,I)${\rm-T}.
Otherwise, it is said $(m,I)${\rm-NPT}.
\end{enumerate}

\end{definition}

\begin{lemma}[Cf. Lemma 3 in \cite{CS09b}]\label{lem:PITRONS}
Fix an interval $I\subset \DR$ and an energy $E\in I$.
Consider an $N$-particle cube $\uC^{(N)}_L(\uu) = \uC^{(n')}_L(\uu') \times \uC^{(n'')}_L(\uu'')$, with  $n',n''\ge 1$,   and a sample of the potential $V(\cdot;\omega)$ such that
\begin{enumerate}[{\rm(a)}]
  \item $\rho\left( \Pi \uC^{(n')}_{L_k}(\uu'), \Pi \uC^{(n'')}_{L_k}(\uu'') \right)  > 2L_k + r_0$; ($r_0$ = the range of interaction)
  \item $\uC^{(N)}_{L_k}(\uu)$ is $E$-non-resonant;
  \item $\uC^{(n')}_{L_{k-1}}(\uu')$ and $\uC^{(n'')}_{L_{k-1}}(\uu'')$ are
  $(m,I)$-NT;
  \item the inductive assumptions $\{\DS{k',n}, \, k'\ge 0, n \le N-1 \}$ hold true.
\end{enumerate}
Then the cube $\uC^{(N)}_{L_k}(\uu)$ is $(E,m)$-non-singular.
\end{lemma}

\begin{lemma}[Cf. Lemma 5 in \cite{CS09b}]\label{lem:NPT.prob.bound}
If $L_0$ is large enough, then
\begin{equation}\label{eq:KFI.bound}
\prob{ \uC_{L_k}(\uu) \text{ is $(m,I)$-PT} } \le \frac{1}{2} L_k^{- p2^{N-(N-1)+1} + 2Nd\alpha}.
\end{equation}
\end{lemma}

Indeed, the "tunneling" property says that one of the projection cubes $\uC^{(n)}_{L_k}(\uu')$ contains at least two distant singular cubes of radius $L_{k-1}$, so that Lemma \ref{lem:NPT.prob.bound} follows  from the inductive assumptions
$\{\DS{k',n}, \, k'\ge 0, 1 \le n \le N-1 \}$ combined with a simple  bound on the number of pairs of points in $\uC^{(n)}_{L_k}(\uu')$ by $\half |\uC^{(n)}_{L_k}(\uu)|^2$.

Introduce the following random variables, depending upon the samples $V(\cdot;\omega)$ :
$$
\begin{array}{ll}
K^{PI}(\uC_{L_{k+1}}(\uu), I) =
&\text{the maximal number of $2NL_k$-distant PI cubes of radius $L_{k}$} \\
&\text{in  $\uC_{L_{k+1}}(\uu)$ which are simultaneously $(E,m)$-S, with $E\in I$,}
\end{array}
$$
$$
\begin{array}{ll}
K^{FI}(\uC_{L_{k+1}}(\uu), I) =
&\text{the maximal number of $2NL_{k}$-distant FI cubes of radius $L_{k}$} \\
&\text{in  $\uC_{L_{k+1}}(\uu)$ which are simultaneously $(E,m)$-S, with $E\in I$.}
\end{array}
$$

The maximal collections of resonant and distant FI (or PI) cubes are, of course, not uniquely defined, but their (maximal) cardinality is well-defined, since there is only a finite number of choices for  cubes $\uC_{L_k}(\uv^{(1)})$, \ldots, $\uC_{L_k}(\uv^{(n)})$ inside  $\uC_{L_{k+1}}(\uu)$.

\begin{cor}\label{cor:2.PI.R}
If $p>Nd\alpha/(2-\alpha)$  and $L_0^{p(2-\alpha) -Nd\alpha}\ge 2$, then
\begin{equation}\label{eq:cor.2.PI.R}
\prob{ K^{PI}(\uC_{L_{k+1}}(\uu), I) \ge 2} \le \frac{1}{2}L_k^{- 4p + 2Nd\alpha}
\le \frac{1}{8}L_{k+1}^{- 2p}.
\end{equation}
\end{cor}

\subsection{Pairs of non-decomposable (FI) cubes}
\label{ssec:FI.pairs}

\begin{lemma}[Cf. \cite{CS09b}]\label{lem:4.FI.boxes}
If $p>2Nd\alpha/(2 -\alpha)$, then
\begin{equation}\label{eq:lem.4.FI}
\prob{ K^{FI}(\uC_{L_{k+1}}(\uu), I) \ge 4} \le \frac{1}{4!}L_{k+1}^{-4p\alpha^{-1} + 4Nd}
\le \frac{1}{4!}L_{k+1}^{-2p}.
\end{equation}
\end{lemma}

For the proof, it suffices to notice that distant FI cubes give rise to \emph{independent} samples of the potential $V$.

As was mentioned before, we see that with sufficiently high probability $K^{PI} + K^{FI} < 3+1=4$, and Lemma \ref{lem:radial.descent.GF} can be used to prove non-singularity of a non-resonant cube of radius $L_{k+1}$ with at most $K^{FI}+K^{PI} \le 4$ pairwise $2NL_k$-distant singular cubes of radius $L_k$ inside it. Indeed, $O(L_k)$-neighborhood of all such cubes can be covered by a collection of annuli of total width $O(L_k)$.

Note that, with $\alpha = 3/2$, it suffices to require that $p> 6Nd$ and $L_0\ge 2$; then the hypotheses on $p$ in Corollary \ref{cor:2.PI.R} and in Lemma \ref{lem:4.FI.boxes} are automatically satisfied.

\begin{cor}\label{cor:2.FI.S}
Assume that the bound \DSk (Eqn \eqref{eq:DS} holds for some $k\ge 0$. Then the bound of the
form \DSkone\, also holds for all $2NL_{k+1}$-distant pairs of FI cubes.
\end{cor}

\proof
Consider two $2NL_{k+1}$-distant FI cubes $\uC_{L_{k+1}}(\ux)$, $\uC_{L_{k+1}}(\uy)$.
Both of them are $(E,m)$-S for some $E\in I$ only if at least one of the following events occurs:
\begin{enumerate}
  \item for some $E\in I$, both $\uC_{L_{k+1}}(\ux)$  and $\uC_{L_{k+1}}(\uy)$ are $E$-R;
  \item $K^{FI}(\uC_{L_{k+1}}(\ux), I) \ge 4$ or $K^{PI}(\uC_{L_{k+1}}(\ux), I) \ge 2$;
  \item $K^{FI}(\uC_{L_{k+1}}(\uy), I) \ge 4$ or $K^{PI}(\uC_{L_{k+1}}(\uy), I) \ge 2$.
\end{enumerate}
By Thm.~\ref{thm:W2.general}, the probability of the event (1) can be bounded by
$\frac{1}{4}L_{k+1}^{-2p}$ (actually, by any power of $L^{-1}_{k+1}$, provided that \eqref{eq:CMnu.stronger} is fulfilled with $B>0$ large enough). In the case where \eqref{eq:CMnu.stronger.Holder} holds, one obtains even a stronger bound by $e^{-L_{k+1}^{\beta'}}$, $\beta'>0$.  Next, it follows from Corollary \ref{cor:2.FI.S} and Corollary \ref{cor:2.PI.R} that the probability of the event (2) (as well of the event (3)) is bounded by
$$
\frac{1}{8}  L_{k+1}^{-2p} + \frac{1}{4!}  L_{k+1}^{-2p} < \frac{1}{4} L_{k+1}^{-2p}.
$$
Therefore, the cubes $\uC_{L_{k+1}}(\ux)$ and $\uC_{L_{k+1}}(\uy)$ are simultaneously $(E,m)$-singular, for some $E\in I$, with probability
$\le \left(\frac{1}{4} + \frac{2}{4} \right) L_{k+1}^{-2p} <  L_{k+1}^{-2p}$, as required.
\qedhere

\subsection{Pairs of decomposable (PI) cubes and mixed pairs}

\begin{lemma}\label{lem:mixed.pairs}
Assume that the bound \DSk (Eqn \eqref{eq:DS} holds for some $k\ge 0$. Then the bound of the
form \DSkone\, also holds for all $2NL_{k+1}$-distant pairs of cubes
$\uC_{L_{k+1}}(\ux)$ and $\uC_{L_{k+1}}(\uy)$, of which at least one is "partially interactive".
\end{lemma}

\proof
Without loss of generality, assume that $\uC_{L_{k+1}}(\uy)$ is PI.
The cubes $\uC_{L_{k+1}}(\ux)$ and $\uC_{L_{k+1}}(\uy)$ are simultaneously $(E,m)$-singular, for some $E\in I$, only if at least one of the following events occurs:
\begin{enumerate}
  \item for some $E\in I$, both $\uC_{L_{k+1}}(\ux)$  and $\uC_{L_{k+1}}(\uy)$ are $E$-R;
  \item $\uC_{L_{k+1}}(\ux)$ is FI, and $K^{FI}(\uC_{L_{k+1}}(\ux), I) \ge 4$ or $K^{PI}(\uC_{L_{k+1}}(\ux), I) \ge 2$;
  \item the PI cube $\uC_{L_{k+1}}(\uy)$ is partially tunneling.
\end{enumerate}
Naturally, the option (2) is to be considered only for a \emph{mixed} pair of cubes: if both of them are PI, it suffices to analyze only the cube $\uC_{L_{k+1}}(\uy)$.

The probabilities of the events (1) and (2) can be estimated exactly as in the proof of Corollary \ref{cor:2.FI.S}, so that their sum does not exceed $\frac{2}{4} L_{k+1}^{-2p}$. By inductive assumptions $\{\DS{k',n},\, k'\ge 0\}$ on systems with $n\le N-1$ particles, the event (3) has probability bounded by $\frac{1}{4} L_{k+1}^{-2p}$. Combining these estimates, the lemma follows.
\qedhere

\medskip

\textit{Now the key bound of the MPMSA, \DSk, is established for all $k\ge 0$. }

\section{From the MSA bounds to multi-particle localization}
\label{sec:MSA.to.DL}

It has been known since more than twenty years that the principal MSA bounds -- in the single-particle theory -- imply the spectral localization; see the original papers \cite{FMSS85,DK89}. In fact, as was pointed out in \cite{CS09a,CS09b}, the main argument used in such a derivation is not specific to single- or multi-particle structure of the random potential.
Note for a reader familiar with the  traditional, single-particle version of Lemma  \ref{lem:MSA.to.SL} given below, that the main argument used in the single-particle context applies to multi-particle systems, with one minor technical modification: instead of \emph{disjoint} pairs of cubes, one should consider \textit{$2NL_k$-distant} pairs of cubes at each scale $L_k$.

In essence, owing to the new eigenvalue concentration bound given by Thm.~\ref{thm:W2.general}, the derivation of 2-particle spectral localization from MSA bounds of the form \DSk\, described in \cite{CS09a} extends -- in a fairly simple way -- to any $N\ge 2$.

\begin{lemma}[Cf. \cite{CS09b}]\label{lem:MSA.to.SL}
Fix an interval $I\subset \DR$ and suppose that the bound \DSk\, (cf. Eqn~\eqref{eq:DS}) holds for all $k\ge 0$. Then with probability one, the Hamiltonian $\uH^{(N)}(\om)$ has pure point spectrum in $I$, and all its eigenfunctions $\uPsi_n(\om)$ with eigenvalues $E_n\in I$ decay exponentially fast at infinity:
$$
\exists\, C_n(\om):\;\forall\, \ux\in\DZ^{Nd}\quad |\uPsi_n(\ux;\om)|
\le C_n(\om) \, e^{-m\| \ux \|}.
$$
\end{lemma}

\proof
As it is well-known, spectrally-a.e. generalized eigenfunction $\uPsi$ of a lattice Schr\"{o}dinger operator $\uH$ is polynomially bounded,
$$
|\uPsi(\ux)| \le Const \, \| \ux \|^A, \; A<\infty,
$$
so it suffices to show that every polynomially bounded solution of the equation $\uH \uPsi = E\uPsi$  with $E\in I$ is exponentially decaying at infinity.

If $\uPsi \not \equiv 0$, then there is a point $\uu\in\DZ^{Nd}$ where $\uPsi(\uu)\ne 0$. We start the analysis of the decay properties of $\uPsi$ by finding the smallest integer $k_0\ge 0$ such that the cube $\uC_{L_{k_0}}(\uzero)$ contains all points $\tau(\uu)$, $\tau\in\fS_N$, obtained by permutations  of the components $u_j$ of the vector $\uu=(u_1, \ldots, u_N)$.

The first observation is that for some $k_1 \ge k_0$ and all $k\ge k_1$ the cubes
$\uC_{L_{k}}(\uzero)$ must $(E,m)$-S. Indeed, assume otherwise. Then there exist an infinite number of values of $k$ (hence, arbitrarily large values of $L_k$) such that $\uC_{L_{k}}(\uzero)$ is $(E,m)$-NS, with $\uu\in \uC_{L_{k-1}}(\uzero)$. Then the non-singularity property implies that for all points  $\ux\in \uC_{L_{k-1}}(\uzero)$, including $\ux=\uu$, one has
$$
\ba
\diy |\uPsi(\ux)| \le e^{-\gamma(m,L_k,N} \, |\partial \uC_{L_{k}}(\uzero) | \,
\max_{ \uy\in \partial^+ \uC_{L_{k}}(\uzero)  } |\uPsi(\uy)| \\
\diy  \le e^{-mL_k} O\left(L_k^{(N-1)d+A}\right) \tto{L_k \to \infty}{0}{}.
\ea
$$
This would mean that $\uPsi(\uu)=0$, which is impossible by our choice of the point $\uu$.

Next, set $\alpha' = 9/8 < \alpha$ and for all $k\ge k_1$ consider the events
$$
\ba
\cE_k =
\left\{ \exists\, \lambda\in I:
\uC_{2L^{\alpha'}_{k+1}}(\uzero)
\text{ contains two $2NL_k$-distant $(\lambda,m)$-S cubes of radius $L_k$}
\right\}
\ea
$$
and
$$
\widetilde{\Omega} := \bigcup_{k\ge k_1} \bigcap_{j\ge k} \left( \Omega \setminus \cE_k \right).
$$
Since $\prob{ \cE_k}  = O(L_k^{2Nd\alpha\alpha' - 2p})$ and $p>3Nd\alpha>Nd\alpha\alpha'$, it follows from Borel--Cantelli lemma that $\prob{\widetilde{\Omega}} = 1$, and for every $\om\in\widetilde{\Omega}$ (i.e., with probability one) there exists $k_2\ge k_1$ such that no cube $\uC_{L_{j+1}}(\uzero)$ with $j\ge k_2$ contains a pair of $(E,m)$-S cubes of radius $L_j$ at distance $\le 2NL_j$.

Further,  introduce a sequence of annuli
$$
\uA_k = \uC_{L^{\alpha'}_{k+1}}(\uzero) \setminus \uC_{L^{\alpha'}_{k}}(\uzero), \; k\ge 0,
$$
and let
$\ux\in \DZ^{Nd} \setminus \uC_{L^{\alpha'}_{k_2}}(\uzero) = \cup_{k\ge k_2} \uA_k$. (Note that it is not necessary to estimate $|\uPsi(\ux)|$ for $\ux\in\uC_{L^{\alpha'}_{k_2}}(\uzero)$ otherwise than by a constant.) Suppose that $\ux\in\uA_j$ for some $j$. Observe that the cubes $\uC_{L_{k}}(\uzero)$ and $\uC_{L_{j}}(\uzero)$  are $2NL_j$-distant and $\om\in\widetilde{\Omega}$, so that one of them is $(E,m)$-NS. Since $k\ge k_2$,
$\uC_{L_{j}}(\uzero)$ is $(E,m)$-S, hence, $\uC_{L_{j}}(\ux)$ is $(E,m)$-NS. Moreover, all cubes of radius $L_k$ outside $\uC_{L_k + 2NL_k}(\uzero)$ are also $(E,m)$-NS. This implies that the function $\uPsi$ is $(L_k,q)$-subharmonic in the cube $\uC_{R}(\ux)$, with
$$
\ba
q = e^{-m\left(L_k + O( \ln L_k)\right)}
\le e^{-m\left(L_k + \frac{1}{2} L_k^{3/4}\right)},
\ea
$$
if $L_k$ is large enough, and
$$
 R := \|\ux\| - (L_k + 2NL_k)), \; \| \ux \| \ge L_k^{9/8} \gg L_k.
$$
It is easy to see that $R/\|\ux\| = 1 - O(L_k^{-1/8})$, and applying Lemma \ref{lem:radial.descent.GF}, we obtain, for $L_k$ large enough,
$$
-  \frac{ \ln |\uPsi(\ux)| } { \| \ux \| } \ge
m\frac{ L_k + L_k^{3/4} } { L_k } \cdot
\left(1 - O\left(L_k^{-1/8}\right)\right)
\ge m,
$$
as required.
\qedhere
\smallskip

Recall that the validity of the hypothesis of Thm.~\ref{thm:MSA.to.DL} is established in Sect.  \ref{ssec:base.MSA}. We come, therefore, to our main result on the $N$-particle spectral localization:

\begin{theorem}\label{thm:SL}
Consider the random operators $\uH^{(N)}(\om)$ of the form \eqref{eq:H}. Suppose that $\uU$ is bounded and $V$ satisfies $\CMnu$. Then there exists $g^*<\infty$ such that if $|g|\ge g^*$, then with probability one the operator $\uH(\om)$ has pure point spectrum, and all its eigenfunctions $\uPsi_n(\om)$ with eigenvalues $E_n\in I$ decay exponentially fast at infinity:
$$
\exists\, C_n(\om):\;\forall\, \ux\in\DZ^{Nd}\quad |\uPsi_n(\ux;\om)|
\le C_n(\om) \, e^{-m\| \ux \|}.
$$
\end{theorem}

\begin{remark}
While the statement of Thm.~\ref{thm:SL} is similar to that of the main result of \cite{CS09b}, a detailed analysis shows that the actual bound on the random constants $C_n(\omega)$ obtained in \cite{CS09b} depends upon the position of the localization center $\ux_n$ for the corresponding eigenfunction $\uPsi_n(\om)$;  Thm.~\ref{thm:W2.general} rules out such a dependence.
\end{remark}

\section{Strong dynamical localization}

To prove the \emph{dynamical} localization, in addition to the assumption \eqref{eq:CMnu.stronger} on the random potential $V(\cdot; \omega)$ we need the following hypothesis: for any finite interval $I\subset \DR$
\begin{equation}\label{eq:Weyl}
\exists \, \kappa = \kappa(I,N,d)<\infty: \; \prob{ \tr ( P_I (\uH_{\uC_{L}(\uu)}(\om)) ) > C L^{\kappa Nd} } \le L^{-B'},
\end{equation}
where $P_I(\cdot)$ is the spectral projection on $I$ and  $B'>0$ will be required below to be sufficiently large, depending on other parameters. It is readily seen that the trace of $P_I(\uH_{\uC_{L}(\uu)})$ grows not faster than linearly in $|\uC_{L}(\uu)|$, if the random potential $V$ is bounded from below. Since we would like to allow also  Gaussian potentials for which $\CMnu$ becomes most simple, we  allow a faster rate of growth. Observe that
$$
\tr( \uH_{\uC_{L}(\uu)}(\omega)) = \tr (\BS{\Delta}_{\uC_{L}(\uu)} + \uU_{\uC_{L}(\uu)})
+ g\, \tr(\uV_{\uC_{L}(\uu)}(\omega)),
$$
and
$$
\tr(\uV_{\uC_{L}(\uu)}(\omega)) = \sum_{\ux\in \uC_{L}(\uu)} \sum_{j=1}^N V(x_j;\omega).
$$
Therefore, for a large class of marginal distributions including Gaussian ones, the required bound follows  from standard results for the sums of IID random variables.

\subsection{Strong dynamical localization}

Here we follow closely the scheme which is well-described in \cite{St01};
see also \cites{GD98,DS01}. Specifically, Propositions 1--6 below correspond to assertions at Steps 1-- 6 from Sect.  3.4 in \cite{St01}.

Introduce the operator $\B{X}$ of multiplication by max-norm in $\ell^2(\DZ^{Nd})$:
$$
(\B{X} \uPsi)(\ux) := \|\ux\|\, \uPsi(\ux), \; \ux\in\DZ^{Nd}.
$$

\begin{theorem}\label{thm:MSA.to.DL}
Assume that the property \DSk\, holds true for a given $N>1$ and for all $k\ge 0$,  with
$p> (3Nd\alpha + \alpha s)/2$, $s>0$. Then the random operators $\uH^{(N)}(\om)$ feature strong dynamical localization in the energy interval $I$: for any finite subset $K\subset \DZ^{Nd}$
and any bounded measurable function $\eta$ with $\supp\, \eta \subset I$,
$$
\esm{  \| \B{X}^s\,  \eta(H(\omega)) \one_K \|  } < \infty.
$$
\end{theorem}

Note that the condition $p>2Nd\alpha/(2-\alpha)$ was required to prove  \DSk.

\smallskip

Since the validity of the hypothesis of Thm.~\ref{thm:MSA.to.DL} is established in Sect.  \ref{ssec:base.MSA}, it implies the $N$-particle dynamical localization for the Hamiltonians $\uH(\om)$:

\begin{theorem}\label{thm:DL}
Assume that the random field $V:\DZ^d\times\Omega\to\DR$ satisfies the condition $\CMnu$ and \footnote{Recall that \eqref{eq:Weyl} is not required for random potentials bounded from below.} \eqref{eq:Weyl} with $B' > 2p - Nd\alpha$. Assume also that the initial scale estimate $\DS{0,N}$ is fulfilled for some interval $I\subset \DR$ with $p> (3Nd\alpha + \alpha s)/2$, $s>0$. Then the random operators $\uH^{(N)}(\om)$ feature strong dynamical localization in the energy interval $I$: for any finite subset $K\subset \DZ^{Nd}$ and any  $\eta\in L^\infty(\DR)$ with
$\supp\, \eta \subset I$,
$$
\esm{  \| \B{X}^s\, \eta(H(\omega)) \one_K \|  } < \infty.
$$
In particular, $\uH^{(N)}(\om)$ features complete dynamical localization with any given value of $s>0$, if  $g$  is large enough: $|g|\ge g^*(s)$, $g^*(s)<\infty$.
\end{theorem}

Now we will describe the strategy of the proof of Thm.~\ref{thm:MSA.to.DL}. Our main goal here is to show that the proof, making use of the simpler and more general eigenvalue concentration bound \eqref{eq:main.W2.bound}, is \emph{very close} to that used in the single-particle context.

\subsubsection{"Bad" and "good" events}
\label{sssec:step1}

Fix $s>0$. We will always assume that
\begin{equation}\label{eq:cond.p}
 p > \max\left\{ 2Nd\alpha/(2 - \alpha),  \,(3Nd\alpha + \alpha s)/2 \right\}
\end{equation}
and set $ b = b(p,N,d,\alpha) :=  2p -2Nd\alpha$. For each $j\ge 1$ consider the events
$$
\begin{array}{l}
\C{S}_j = \{ \omega:\, \exists\, E\in I\,\,  \exists\, \B{y},\B{z}\in \uC_{4(N+1)L_{j+1}}(\uzero) \text{ such that } d_S(\B{y}, \B{z})>2NL_{j}\\
\;\;\;\;\;\;\;\;\;  \text{ and }\uC_{L_{j}}(\B{y}), \uC_{L_{j}}(\B{z}) \text{ are  $(m,E)$-singular} \}.
\end{array}
$$
and (in the case where the random potential $V$ is \emph{not} bounded from below)
$$
\begin{array}{l}
\C{T}_j = \{ \omega:\, \tr ( P_I (\uH_{\uC_{L_{j+2}}(\uu)}) ) > C L_{j+2}^{\kappa Nd} \;\}.
\end{array}
$$
Further, for $k\ge 1$ denote
$$
\Omega_{k}^{(bad)} = \bigcup_{j\ge k} \left( \C{S}_j \cup \C{T}_j \right)
$$
and consider the annuli
$$
\B{M}_k = \uC_{4(N+1)L_{k+1}}(\uzero) \setminus \uC_{4(N+1)L_{k}}(\uzero).
$$

\begin{prop}\label{prop:step1}
Under the assumption \eqref{eq:Weyl} with $B' >  2p-2Nd\alpha$
$$
\forall\, k\ge 1\quad \prob{ \Omega_{k}^{(bad)} }
\le c(\alpha,d, p,N ) L_k^{ -(2p - 2Nd\alpha) }.
$$
\end{prop}

\proof
The  number of pairs  $\B{y},\B{z}\in \uC_{4(N+1)L_{j+1}}(\uzero)$ figuring in the definition of the event $\C{S}_j$ is bounded by
$\half (4(N+1)L_{j+1})^{2Nd} = C(N,d) L_{j}^{2Nd\alpha}$, and $\DS{j,N}$ says that
$$
\prob{ \uC_{L_{j}}(\B{y}), \uC_{L_{j}}(\B{z}) \text{ are  $(m,E)$-singular} } \le L_{j}^{-2p},
$$
while  for the event $\C{T}_j$ we have the bound \eqref{eq:Weyl}. Now we require that the exponent $B'$ be large enough, so that $\prob{\C{T}_j} \le  L_j^{-2p+2Nd\alpha}$ and
\begin{equation}\label{eq:proof.prop.1}
\begin{array}{l}
\displaystyle \prob{ \Omega_{k}^{(bad)} } \le \sum_{j\ge k} \left(\prob{\C{S}_j} +\prob{\C{T}_j} \right)
\displaystyle \le  \sum_{j\ge k} Const L_{j}^{2Nd\alpha} \cdot 2\, L_j^{-2p} \\
\displaystyle = L_k^{-b} \left[ 1 + \sum_{j > k} L_k^{b} L_k^{-b\alpha^{j-k}} \right]
\le Const \, L_k^{-(2p - 2Nd\alpha)}.
\end{array}
\end{equation}
\qedhere

\subsubsection{Centers of localization}
\label{sssec:step2}
By Thm.~\ref{thm:SL}, there exists a subset $\Omega_1\subset\Omega$ with $\prob{\Omega_1}=1$ such that for any $\omega\in\Omega_1$ the spectrum of $\uH^{(N)}(\omega)$ in $I$ is pure point. Fix $\omega\in\Omega_1$ and let $\uPhi_n(\omega)$ be a normalized eigenfunction of $\B{H}(\omega)$,  with  eigenvalue $E_n(\omega)\in I$.
We call a \textit{center of localization} for $\uPhi_n$ every point $\B{x}_n(\omega)\in\D{Z}^{Nd}$ such that
\begin{equation}\label{eq:center.loc}
|  \uPhi_n(\B{x}_n(\omega))  | = \max_{\B{y}\in\D{Z}^{Nd} } \, | \uPhi_n(\B{y}) |.
\end{equation}
Since $\uPhi_n\in \ell^2(\D{Z}^{Nd})$, such centers always exist and, due to the normalization $\|\uPhi_n\|=1$, the number of centers of localization $\B{x}_{n,a}$ for a given $n$ must be finite.

\begin{prop}\label{prop:step2}
There exists $k_0$ such that for all $\omega\in\Omega_1$ and $k\ge k_0$, if
one of the centers of localization $\B{x}_{n,a}$ for an eigenfunction $\uPsi_n$ with eigenvalue $E_n\in I$ belongs to  a cube $\uC_{L_k}(\B{x})$, then the cube
$\uC_{L_{k+1} }(\B{x})$ is $(m,E_n)$-S.
\end{prop}

\proof
Assume otherwise. Then the GRI for the eigenfunctions \eqref{eq:GRI.EF} combined with the non-singularity condition \eqref{eq:def.NS} implies that, for $L_k$ large enough,
$$
\begin{array}{l}
\displaystyle | \uPsi_n(\ux_{n,a})| \le e^{-\gamma(m,L_{k+1},N)} \, |\partial \uC_{L_{k+1}}(\ux)|
\max_{\uy \in \partial^+ \uC_{L_{k+1}}(\ux)} | \uPsi_n(\uy) |\\
\displaystyle <  \max_{\uy \in \partial^+ \uC_{L_{k+1}}(\ux)} | \uPsi_n(\uy)|,
\end{array}
$$
which contradicts the definition of a center of localization.
\qedhere

We will denote by  $\B{x}_{n,1}$ is the center of localization closest to the origin; again, it might be not unique, but this is irrelevant for the proofs.
%
Fix $k_0$ as in  Prop. \ref{prop:step2} and set
$$
\Omega_{k}^{(good)} =  \Omega_1 \setminus \Omega_{k}^{(bad)}.
$$

\begin{prop}\label{prop:step3}
There exists $j_0 = j_0(m,\alpha,d)$ large enough such that for $j\ge j_0$, $j\ge k$ and $\B{x}_{n,1}\in\uC_{L_{j+1}}(\uzero)$
$$
\begin{array}{l}
\| (1 - \one_{\uC_{4(N+1)L_{j+2}}(\uzero)})\,  \uPhi_n\| < \frac{1}{4}.
\end{array}
$$
\end{prop}

\proof
Using the annuli $\B{M}_k$, we can write
$$
\begin{array}{l}
\displaystyle \| (1 - \one_{\uC_{4(N+1)L_{j+1}}(\uzero)})\,  \uPhi_n \|^2 =
 \sum_{i \ge j+2} \| \one_{\B{M}_i} \uPhi_n \|^2
\displaystyle = \sum_{i \ge j+2} \sum_{ \uy\in \B{M}_i} |  \uPhi_n(\uy) |^2 \\
\end{array}
$$
Fix $i\ge j+2$. The cube $\uC_{L_{i-1}}(\uzero) \supseteq \uC_{L_{j+1}}(\uzero)$ contains the center of localization  $\B{x}_{n,1}$, so, by Prop. \ref{prop:step2}, $\uC_{L_{i}}(\uzero)$ is $(E_n,m)$-singular. By construction of the event $\Omega_{k}^{(good)}$, $k\le j$,  the cube $\uC_{L_{i}}(\uy)$
must be $(E_n,m)$-NS. In turn, this implies by the GRI for eigenfunctions that
$|\uPhi_n(\uy) |^2 \le e^{-2mL_{i}}$. Now the claim follows from a polynomial (in $L_{i}$) bound on the number of terms in the sum $\sum_{ \uy\in \B{M}_i}(\cdot)$.
\qedhere


\begin{prop}\label{prop:step4}
There exists $c_4 = c_4(m,d,\kappa)$ such that for $\omega\in\Omega_{k}^{(good)}$ and
$j\ge k$ the following bound holds:
\begin{equation}\label{eq:DD.7}
\card \left\{ n:\, \B{x}_n\in \uC_{L_{j+1}}(\uzero) \right\} \le c_4\, L_{j+1}^{\alpha \kappa d}.
\end{equation}
\end{prop}

\proof
Observe that we have
\begin{equation}\label{}
\begin{array}{l}
\displaystyle
\sum_{\ux_{n,1} \in \uC_{L_{j+1}}(\uzero)}
\left( \one_{\uC_{L_{j+2}}(\uzero) } P_I( \uH) \one_{\uC_{L_{j+2}}(\uzero)}  \uPhi_n, \, \uPhi_n  \right) \le \tr\, (  \one_{\uC_{L_{j+2}}(\uzero) } P_I( \uH) ).
\end{array}
\end{equation}
It suffices to show that each term in the LHS is bigger than $\half$. Indeed, by Prop. \ref{prop:step3},
$$
\begin{array}{l}
\left( \one_{\uC_{L_{j+2}}(\uzero)} P_I( \uH) \one_{\uC_{L_{j+2}}(\uzero)} \,\uPhi_n,\,\uPhi_n \right) \\
= \left( \one_{\uC_{L_{j+2}}(\uzero)} P_I( \uH)  \,\uPhi_n,\,\uPhi_n \right)
- \left( \one_{\uC_{L_{j+2}}(\uzero)} P_I( \uH) (1 - \one_{\uC_{L_{j+2}}(\uzero)}) \,\uPhi_n,\,\uPhi_n \right) \\
\ge  \left( \one_{\uC_{L_{j+2}}(\uzero)} \,\uPhi_n,\,\uPhi_n \right) - \frac{1}{4}
=  \left( \uPhi_n,\,\uPhi_n \right)
-  \left( (1 - \one_{\uC_{L_{j+2}}(\uzero)}) \,\uPhi_n,\,\uPhi_n \right) - \frac{1}{4} \\
\ge \half.
\end{array}
$$
\qedhere

\subsubsection{Eigenfunction correlator bounds }
\label{sssec:step5}

\begin{prop}\label{prop:step5}
 There exists an integer $k_1 = k_1(\kappa, L_0)$ such that for all $k\ge k_1$,
$\omega\in\Omega_{k}^{(good)}$ and $\B{x}\in \B{M}_{k}$
\begin{equation}\label{eq:DD.11}
| \eta(H(\omega))(\ux, \uzero) |
\le e^{-m L_{k-1}/2} \| \eta\|_\infty.
\end{equation}
\end{prop}

\proof
Without loss of generality, assume that $\|\eta\|_\infty \ne 0$:
\begin{equation}\label{eq:prop5.1}
\begin{array}{l}
\displaystyle \| \eta\|_\infty|^{-1} \eta(H(\omega))(\ux, \uzero) |
\displaystyle \le \sum_{n:\,E_n\in I} \| \eta\|_\infty|^{-1} |\eta(E_n)|\, | \uPhi_n(\ux) |  | \uPhi_n(\uzero)|
\\
\displaystyle \le  \sum_{\substack{n:\,E_n\in I \\ \ux_{n,1}\in \B{M}_{k}(\uzero) }} \,
 | \uPhi_n(\ux) | \, | \uPhi_n(\uzero)|
\displaystyle  + \sum_{j > k}  \sum_{\substack{n:\,E_n\in I \\ \ux_{n,1}\in \B{M}_{j}(\uzero) }} \,
 |\uPhi_n(\ux) | \, | \uPhi_n(\uzero)|.
 \end{array}
\end{equation}
For $k\ge k_0$ and $L_k$ large enough, the first sum in the RHS can be bounded as follows:
\begin{equation}\label{eq:prop5.3}
\begin{array}{l}
\displaystyle \sum_{\substack{n:\,E_n\in I \\ \ux_{n,1}\in \B{M}_{k}(\uzero) }} \,
 | \uPhi_n(\ux) | \, | \uPhi_n(\uzero)|
  \le const\,  L_{k+1}^{\alpha \kappa Nd} \, e^{-mL_k} \le \half e^{-mL_k/2},
 \end{array}
\end{equation}
since  one of the cubes $\uC_{L_{k}}(\ux)$, $\uC_{L_{k}}(\uzero)$ must be $(E_n,m)$-NS: indeed, these cubes are $2NL_{k}$-distant. Next,  fix any $j\ge k+1$ and consider the sum with $\ux_{n,1}\in \B{M}_{j}(\uzero)$. The cubes $\uC_{L_{j}}(\uzero)$ and $\uC_{L_{j}}(\ux_{n,1})$ are $2NL_{j}$-distant and by Prop. \ref{prop:step2},
for $k$ (hence, $j$)  large enough
the cube $\uC_{L_{j}}(\ux_{n,1})$ is  $(E_n,m)$-S, so for $\omega\in\Omega_{(k)}^{good}$ the cube  $\uC_{L_{j}}(\uzero)$ must be $(E_n,m)$-NS. Therefore,
$$
| \uPhi_{n} (\uzero)| \le Const \, e^{-mL_j}.
$$
Using again Prop. \ref{prop:step4}, we get, for $L_k$ large enough,
\begin{equation}\label{eq:prop5.4}
\sum_{j > k}  \sum_{\substack{n:\,E_n\in I \\ \ux_{n,1}\in \B{M}_{j}(\uzero) }} \,
 | \uPhi_n(\ux) |  | \uPhi_n(\uzero)|
  \le Const \sum_{j>k} e^{-mL_j} L_j^{\alpha \kappa Nd}
 \le \half e^{-mL_k/2}.
\end{equation}
Now the claim follows from \eqref{eq:prop5.3} and \eqref{eq:prop5.4}.
\qedhere

\begin{prop}\label{prop:step6}
Fix $k_1$ as in Prop. \ref{prop:step5}. Then for any $k\ge k_1$ and $\B{x}\in\B{M}_k$,
\begin{equation}\label{eq:prop6}
\esm{ \| \one_{\uC_{L_k}(\B{x})} \eta(H(\omega)) \, \one_{\uC_{L_k}(\uzero)}  \| }
\le \| \eta \|_\infty
\left( C L_k^{-2p  + 2Nd\alpha} + e^{-m L_k/2} \right).
\end{equation}
\end{prop}

\proof
Using   Prop. \ref{prop:step5} and Prop. \ref{prop:step1}, we can write
$$
\begin{array}{l}
\esm{ \| \one_{\uC_{L_k}(\B{x})} \eta(H(\omega)) \, \one_{\uC_{L_k}(\uzero)}  \| } \\
= \esm{ \one_{ \Omega_k^{(bad)} }\,  \| \one_{\uC_{L_k}(\B{x})} \eta(H(\omega)) \, \one_{\uC_{L_k}(\uzero)}  \| }
+ \esm{ \one_{ \Omega_k^{(good)} }\,  \| \one_{\uC_{L_k}(\B{x})} \eta(H(\omega)) \, \one_{\uC_{L_k}(\uzero)}  \| } \\
\le \| \eta \|_\infty \left(\prob{ \Omega_k^{(bad)}} + e^{-mL_k/2} \right)
\le \| \eta \|_\infty \left( C L_k^{-2p  + 2Nd\alpha}  + e^{-mL_k/2} \right).
\end{array}
$$
\qedhere

\subsubsection{Conclusion}
\label{sssec:step7}

Fix a set $K \subset \D{Z}^{Nd}$ and find $k\ge k_1$ such that
$K\subset \uC_{L_k}(\uzero)$. Then
$$
\ba
\diy \esm{  \| \B{X}^s\, \eta(H(\omega)) \one_K \|  }
\le c_{Nd} L_k^s
+ \sum_{j\ge k} \D{E} \left[   \| \B{X}^s \, \one_{\B{M}_{j} } \eta(H(\omega)) \one_K \|  \right]
\\
\diy \le c(k) +
\sum_{j\ge k} c_{Nd} L_{j+1}^s
\left(
\sum_{ \B{w} \in \B{M}_j }
\D{E} \left[  \| \one_{\uC_{L_k}(\B{w})} \eta(H(\omega))
\one_{\uC_{L_k}(\uzero)} \|  \right]
\right)
\\
\diy \le C\left[  1 + \sum_{j\ge k} L_j^{\alpha s} L_j^{Nd\alpha}
\left( L_j^{-2p + 2Nd\alpha } + e^{-m L_j/2} \right)
\right] < \infty,
\ea
$$
since $2p -3Nd\alpha - \alpha s >0$,  and $L_j=(L_0)^{\alpha^j}$ grow fast enough.

\medskip

This completes the proof of Thm.~\ref{thm:MSA.to.DL}.

\section{Appendix. Proof of the Radial Descent bound (Lemma \ref{lem:radial.descent})}
\label{sec:RDL.proof}
\def\cSt{{\widetilde{\cS}}}
\def\cRt{{\widetilde{\cR}}}
\proof
It suffices to consider non-negative functions; otherwise, we replace $f$ by $|f|$.

Let a function $f:C_L(0)\to\DR_+$ be $(\ell,q,\cS,c)$-subharmonic. Introduce "spheres"
$S_r = \{x:\, \|x\|=r\}$ and the sets
$$
\cS' = \{x:\, S_{\|x\|} \cap \cS \ne \varnothing \},
\quad \cS'' = \bigcup_{r:\, S_r \cap \cS \ne \varnothing  }
\; \bigcup_{j=0}^{c\ell} S_{r+j}
$$
and also
$$
 \cR = \{ r\ge 0:\, C_r(0) \subset C_L(0)\setminus \cS'' \}.
$$
Note that if $\|x\| = r\in\cR$, then
$$
f(x) \le q \max_{ y:\, \|y-x\|\le \ell} f(y)
\le q \max_{ y:\, \|y\|\le r+\ell} f(y).
$$
Moreover, for any $u$ with $\|u\|\in [r-c\ell, r]$ the definition of the sets $\cR$ and $\cS''$ implies that $u\not\in\cS$, so that
\be\label{eq:monot.1}
\ba
\diy
f(u) \le q \max_{ y:\, \|y-u\|\le \ell} f(y)
\le q \max_{ y:\, \|y\|\le \|u\|+\ell} f(y) \\
\diy \le q \max_{ y:\, \|y\|\le r+\ell} f(y).
\ea
\ee
Further, for any $u$ with $\|u\| <r-c\ell$ there exists a value
\footnote{$R=(1+c)\ell$ if $u\in\cS$, and $R=\ell$, otherwise.}
$R\in \{\ell, (1+c)\ell\}$ such that
\be\label{eq:monot.2}
\ba
\diy f(u) \le q \max_{ y:\, \|y-u\|\le R} f(y)
\le q \max_{ y:\, \|y\|\le (r-c\ell)+(1+c)\ell} f(y) \\
\diy \le q \max_{ y:\, \|y\|\le r+\ell} f(y).
\ea
\ee
Combining \eqref{eq:monot.1} with \eqref{eq:monot.2}, we conclude that for any $r\in\cR$
\be\label{eq:monot.3}
\max_{u\in C_{r}(0)} f(u) \le q \max_{y\in C_{r+\ell}(0)} f(y).
\ee
Construct a sequence of points $\{r_n\ge 0, 0 \le n \le M\}$ by recursion:
$$
r_0 = L;
\; r_{n}  = \max \{ r\in \cR:\,  r \le r_{n-1} - \ell  \}, 1\le n \le M,
$$
with some $M = M(\cR)$. Set, formally, $r_{M+1} = 0$. We see that, as long as $r_n\ge 0$, either $r_n = r_{n-1} - \ell$, or $J_n := [r_n, r_{n-1}-\ell] \subset \cR^c$. The total length of all non-empty intervals $J_n$ is bounded by the total width $W(\cS'')$ of annuli covering $\cS''$. Now we can write
$$
\ba
\diy L = \sum_{n=0}^M (r_{n-1} - r_{n}) \\
\diy = (r_{M} - r_{M+1}) + \sum_{n: J_n \ne \varnothing} (r_{n-1} - r_{n})
+ \ell \card\{n: r_n = r_{n-1} - \ell\}  \\
\le \ell+ W(\cS'') + \ell \card\{n: r_n = r_{n-1} - \ell\}
\ea
$$
yielding
$$
M \ge \card\{n: r_n = r_{n-1} - \ell\}
\ge \frac{L - W(\cS'')}{ \ell } - 1 \ge \frac{L - W_{\cA}}{ \ell } - 1.
$$
Further, $W(\cS'') \le W_{\cA}$, since the annuli $A_i\in\cA$, by assumption, cover the $c\ell$-neighborhood of the set $\cS$. Therefore,
$$
f(0) \le q^M \|f\|_\infty.
$$
More generally, if we stop the construction of the sequence $\{r_n\}$ at $n=M'$ as soon as
$r_{M'+1} < r$, for some given $r > W_{\cA} +\ell \ge W(\cS'') + \ell$, then we obtain
$$
f(r) \le q^{M'} \|f\|_\infty,
\quad
M' \ge \frac{L - r - W_{\cA}}{ \ell } - 1.
$$
This completes the proof of Lemma \ref{lem:radial.descent}.
\qedhere


\end{document}